\documentclass{svproc}
\usepackage{url}

\usepackage{amssymb}
\usepackage{multicol} % for multicolumn list
\usepackage{tabularx} % ... for the same ... \begin{table}[h] \begin{tabularx}{\textwidth}{|X|X|} a & b \\ c & d \\ \end{tabularx} \end{table}
\usepackage[normalem]{ulem} % to use strike out \sout{...}

\newcommand{\op}{\begin{itemize}}
\newcommand{\ed}{\end{itemize}}
\newcommand{\opp}{\begin{quote}}
\newcommand{\edd}{\end{quote}}
\newcommand{\im}{\item[$\bullet$]}
\newcommand{\xm}{\item[]}

\newcommand{\yes}{\mathtt{Yes}}
\newcommand{\no}{\mathtt{No}}

\newcommand{\PPP}{(\HH, \EE, \WW, \LL)}
\newcommand{\HH}{\mathsf{H}}
\newcommand{\EE}{\mathsf{E}}
\newcommand{\WW}{\mathsf{W}}

\newcommand{\LL}{\mathsf{Loss}}
\newcommand{\PP}{\mathbb{P}}
\newcommand{\NN}{\mathbb{N}}

\newcommand{\XX}{\mathsf{X}}
\newcommand{\YY}{\mathsf{Y}}

\newcommand{\data}{e_1 \cdots e_n}
\newcommand{\stream}{e_1 e_2 \cdots}
\newcommand{\duno}{\texttt{?}}
\begin{document}
\mainmatter              % start of a contribution
\title{Unified Inductive Logic: From Formal Learning to Statistical Inference to Supervised Learning}
\titlerunning{Unified Inductive Logic}  % abbreviated title (for running head)
\author{Hanti Lin
}
\authorrunning{Hanti Lin} % abbreviated author list (for running head)
\tocauthor{Ivar Ekeland, Roger Temam, Jeffrey Dean, David Grove,
Craig Chambers, Kim B. Bruce, and Elisa Bertino}
\institute{University of California, Davis, CA 95616, USA\\
\email{ika@ucdavis.edu}
}

\maketitle              % typeset the header of the contribution
\begin{abstract} While the traditional conception of inductive logic is Carnapian, I develop a Peircean alternative and use it to unify formal learning theory, statistics, and a significant part of machine learning: supervised learning. Some crucial standards for evaluating non-deductive inferences have been assumed separately in those areas, but can actually be justified by a unifying principle.

% The abstract should briefly summarize the contents of the paper in 150--250 words.

\end{abstract}

\section{Introduction}

According to the Carnapian/Bayesian view, inductive logic  is a matter of securing a ``high proportion'' of worlds that make the conclusion true within the domain of the worlds that make the premises true (Carnap 1945). An alternative view of inductive logic is developed here: it is a matter of being guaranteed to get a true conclusion given ``enough'' premises (or data, or evidence).
%%% deductive logic is the limiting case that trivializes the concept of ``enough''. 

This idea can be traced back to C. S. Peirce (1902/[1994]: CP 2.780-1), and I propose that it be systematically developed along the following line of thought: We should be given a certain kind of guarantee {\em at least} when the amount of evidence is arbitrarily large. What kind of guarantee? A natural idea is to seek (i) a guarantee to {\em actually} get exactly the true answer to the question posed in one's context of inquiry. If that is unachievable, we should seek (ii) a guarantee to have a {\em high physical objective probability} of getting exactly the true answer. If even that is unachievable, we should seek, or settle with, (iii) a guarantee to have a high probability of getting {\em close} to the true answer. Note that those three guarantees, (i)-(iii), form a sequence of increasingly lower standards:
\opp 
\op 
\item[(i)] a guarantee to \emph{\uline{actually}} get exactly the true answer,
\xm $\quad\quad\quad\quad\quad\quad\quad\; \downarrow$ \footnotesize{\sf weaken}
\item[(ii)] a guarantee to \emph{\uline{have a high chance}} of getting \emph{\uline{exactly}} the true answer,
\xm $\quad\quad\quad\quad\quad\quad\quad\quad\quad\quad\quad\quad\quad\quad\quad\quad\quad\quad\quad\quad\;\; \downarrow$  \footnotesize{\sf weaken again}
\item[(iii)] a guarantee to have a high chance of getting \emph{\uline{close to}} the true answer.
\ed 
\edd 
When we tackle an empirical problem, we ought to strive for the highest achievable of such standards---achievable relative to the empirical problem undertaken. In a slogan: {\em strive for the highest achievable!} This is a principle that, I claim, unifies multiple areas that study non-deductive inferences. In particular, many disciplines---including formal learning theory, statistical testing theory, statistical estimation theory, and supervised learning in machine learning---appear to {\em assume} distinct standards to evaluate the different types of inference methods that they each study. But, if I am right, those evaluative standards need not be {\em assumed}, let alone assumed {\em separately}; instead, they can be justified by the unifying principle I propose. 

To make my claim precise, many familiar concepts (such as empirical problems and various modes of convergences) need to be reformulated in a uniform setting. This will be addressed in the first half of this paper (through section 3). Then, in the second half (beginning in section 4), the main mathematical results can be stated, and their philosophical significance will be explained. I will conclude by looking into both the history and the future: revisiting Peirce's original ideas and exploring the potential for further unification, including a version of Bayesianism. 

\section{Varieties of Empirical Problems}

Examples first:

\begin{example}[Enumerative Induction]
	The {\bf easy raven problem} poses a question: Are all ravens black? Two potential answers: $\yes$ vs. $\no$, which form the hypothesis space $\HH = \{\yes, \no\}$. Evidence is gathered by observing ravens one by one, and noting their colors as either black (\texttt{1}) or nonblack (\texttt{0}). So the space of the possible evidential states, $\EE$, is the tree of all finite binary sequences. A possible world for this problem takes the form: $$w \;=\; (e_1 e_2 e_3 \cdots, h),$$
	where $e_1 e_2 e_3 \cdots$ is an infinite binary sequence, and $h$ is the competing hypothesis true in that world $w$. It is assumed in the background that either all ravens are black or, if not, a counterexample would be observed sooner or later if the evidence were to accumulate indefinitely. This assumption rules out only one possible world, $(\texttt{111}\cdots, \no)$, in which the true answer is $\no$ (not all ravens are black) and we would still always only observe black ravens $\texttt{111}\cdots$. This background assumption is formalized by a set $\WW$ of possible worlds---the set of all worlds of the form $(e_1 e_2 e_3 \cdots, h)$ except for $(\texttt{111}\cdots, \no)$. %To finish off, let $\LL(h, w)$ be $1$ minus the truth value of $h$ in $w$. ... let $\LL(h, w) = 0$ if $h$ occurs in $w$ (i.e., if $h$ is true in $w$); otherwise $\LL(h, w) = 1$. 
\end{example}

If the above background assumption is relaxed to include the possible world $(\texttt{111}\cdots, \no)$, we obtain the {\em hard} raven problem, which was studied in formal learning theory only quite recently (Lin 2022).

The above example suggests that, in general, an empirical problem has at least three components: (i) competing hypotheses, (ii) data sequences as possible evidence, (iii) a background assumption. Indeed, these three components also figure in another classic empirical problem:

\begin{example}[Statistical Testing]
	The {\bf fair coin problem} poses a question: Is the coin fair? So the hypothesis space $\HH$ is $\{\texttt{Fair},\texttt{Unfair}\}$. Evidence is obtained by tossing the coin repeatedly, and observing the results, either landing heads (\texttt{1}) or landing tails (\texttt{0}). So the evidence space $\EE$ for this problem is the tree of binary sequences (as in the easy raven problem). Assumed in the background is the standard IID assumption in statistics: that the bias $\theta$ of the coin, i.e., the probability of landing heads, stays constant through time and coin tosses are independent. Under the IID assumption, each possible bias $\theta$ in the unit interval $[0, 1]$ determines a probability function $\PP_\theta$ defined over $\EE$ (which is the binary tree). A possible world for this problem takes this form: 
	$$w \;=\; (e_1 e_2 e_3 \cdots, \theta, \PP_\theta).$$ 
	In this world, $\theta$ is the true bias of the coin, $\PP_\theta$ is the true probability function that represents the data-generation mechanism, and it turns out to generate the data sequence $e_1 e_2 e_3 \cdots$. The background assumption is represented by a set of possible worlds, $\WW$, defined as the set of all worlds of the above form. The hypothesis $\texttt{Fair}$ asserts that $\theta = 0.5$, which is true in the worlds in which $\theta = 0.5$. Similarly for the hypothesis $\texttt{Unfair}$, which asserts the negation $\theta \neq 0.5$.  %To define the loss function: $\LL(\texttt{Fair}, w) = 0$ if the (true) bias $\theta_w$ in $w$ is $.5$; otherwise $\LL(\texttt{Fair}, w) = 1$. $\LL(\texttt{Unfair}, w) \,=\, 1 - \LL(\texttt{Fair}, w)$.	
\end{example}

Some clarifications are in order. First, the background assumption is very weak. For example, it is logically compatible with this possible world:
$$w \;=\; (\texttt{1010} \cdots, 0.5, \PP_{0.5}),$$
in which the coin is fair and alternates between landing heads ($\texttt{1}$) and tails ($\texttt{0}$). It is also logically compatible with this possible world:
$$w \;=\; (\texttt{1111} \cdots, 0.5, \PP_{0.5}),$$
in which the coin is fair and it turns out to always land heads---yes, this is logically possible. In fact, the background assumption is even compatible with any worlds of the following form, where $e_1 e_2 e_3 \cdots$ is an arbitrary binary sequence:
$$w \;=\; (e_1 e_2 e_3 \cdots, 0.5, \PP_{0.5}),$$
in which the coin is fair and it turns out to land heads or tails according to the pattern $e_1 e_2 e_3 \cdots$. All those worlds are ruled in, following the practice of classical statistics.

Second, note that probabilities are assigned to the data sequences in the evidence space. That is, each probability function $\PP_\theta$ is defined over the evidence space, which represents a stochastic mechanism for generating data or evidence and conforms to the use of {\em objective physical probabilities}, or {\em chances}, in classical statistics. Bayesian statistics, and Bayesian epistemology in general, allow probabilities to be assigned to possible worlds, but those probabilities represent subjective degrees of belief. A comparison of the present work with Bayesianism will be provided in the concluding section. 

The fair coin problem presupposes that there exist probabilities, so probabilities have to figure in the possible worlds in use. In contrast, for the easy raven problem, it suffices to use worlds of simpler forms without probabilities. So there should be flexibility in our designs of possible worlds:

\begin{definition}[Possible World]
A possible world, or world for short, is an ordered pair or tuple of the form: $$w \;=\; (e_1 e_ 2 \cdots, \mathsf{other}),$$ where the first component $e_1 e_ 2 \cdots$ is an infinite data sequence, understood as the one produced in that world, and the second component $\mathsf{other}$ specifies the other relevant elements of that world.
\end{definition}

Many different things can go into $\mathsf{other}$, such as a distinguished statement true in world $w$, or the distribution of physical probabilities that exist objectively in that world. 

I have mentioned three components of an empirical problem: (hypothesis, evidence, and assumption). There is a fourth component, to be motivated below:

\begin{example}[Statistical Estimation]
	The {\bf coin bias problem} is basically the same as the fair coin problem except that it poses a more fine-grained question: What is the bias of the coin? So the hypothesis space is $\HH = [0, 1]$, the unit interval. The evidence space $\EE$ is the same; so is the background assumption $\WW$. The possible worlds in use are the same, too, taking the form $w = (e_1 e_2 e_3 \cdots, \theta, \PP_\theta)$. But our guess of the bias, as a hypothesis $h \in \HH$, can be more or less accurate; the loss of accuracy can be measured by the difference between the guess $h$ and the true bias. So the loss of accuracy of guessing $h \in [0, 1]$ in a world $w$, written $\LL(h, w)$, can be defined as the difference between the guess $h$ and the true bias in $w$. 
\end{example}

So we need this additional definition:

\begin{definition}[Loss Function]
A loss function is a function $\LL: \HH\times\WW \to \mathbb{R}$ that maps any hypothesis $h$ in $\HH$ together with any world in $\WW$ to a nonnegative real number, denoted by $\LL(h, w)$, under the constraint that, in each world $w \in \WW$, there is a unique hypothesis $h \in \HH$ that attains zero loss of accuracy: $\LL(h, w) = 0$ .
\end{definition}

The uniqueness constraint is adopted in this paper only for the sake of simplicity. It allows us to talk about convergence to {\em the truth} in a world $w$: the unique hypothesis that attains zero loss of accuracy in that world $w$. It is not hard to generalize, referring to \emph{a} truth or truth\emph{s} in a world, but let's opt for simplicity here in order to focus on more important issues.

%\noindent {\em Examples 1 and 2 Revisited.} 
When we were thinking about the first two examples (the easy raven problem and the fair coin problem), we were not compelled to think about the loss function only because the role of the loss function is already played by the talk of truth and falsity. That is, the loss of accuracy has only two values: 0 for getting the truth, 1 for getting a falsehood. Or more precisely, in the easy raven and fair coin problems, the loss function is simple: $\LL(h, w)$ is equal to $1$ minus the truth value of hypothesis $h$ in world $w$.

Thus, for the sake of uniformity, every empirical problem is required to include a loss function as a fourth component. In fact, as will become clear in the next section, this fourth component is even a necessity in order to produce a uniform treatment of various standards for evaluating inference methods. Hence the following definition:

\begin{definition}[Empirical Problem]
An empirical problem, or problem for short, is a quadruple $(\HH, \EE, \WW, \LL)$ with the following interpretations and requirements:
	\op 
	\im $\HH$ is a set of hypotheses, understood as the hypotheses under consideration.
	
	\im $\EE$ is a tree, called the evidence tree, in which the nodes are finite data sequences $e_1 \cdots e_n$ ordered by sequence extension, and every branch is infinite.
	
	\im $\WW$ is a set of possible worlds such that the infinite data sequences that appear in the worlds therein are exactly the branches of the evidence tree $\EE$. This set $\WW$ is meant to contain all and only the possible worlds compatible with one's background assumption.
	
	\im $\LL$ is a loss function on $\HH \times \WW$.
	\ed 	
\end{definition}

This definition is also general enough to cover problems in supervised learning, from binary classification to nonparametric regression, to be discussed in section \ref{sec-supervised}. 

The components of an empirical problem have their own roles to play. The evidence space $\EE$ and the hypothesis space are used to define inference methods as functions from the former to the latter (with a minor refinement to be formally stated below). The other two components are used to define various standards for assessing inference methods, as we will see below. Those standards examine each inference method by considering how that method performs for truth seeking across a range of the possible worlds---the worlds in $\WW$. So, $\WW$ represents the background assumption against which inference methods are evaluated. These ideas will be fleshed out in the formal definitions provided in the next section.

\section{Modes of Convergence as Evaluative Standards}\label{sec-modes} 

Here are the objects of evaluation:

\begin{definition}[Inference Method]
	An inference method for an empirical problem $\PPP$ is a function $M: \EE \to \HH \cup \{\texttt{\em ?}\}$; that is, $M$ can receive any finite data sequence that figures as a node of the evidence tree $\EE$, and then output one of the hypotheses in $\HH$ or a question mark $\texttt{\em ?}$ that represents judgment suspension. 
\end{definition}

Only the first two components of a problem, $\HH$ and $\EE$, are needed to define the inference methods for that problem. The remaining two components, $\WW$ and $\LL$, are used to define standards for assessing inference methods. But preliminaries first:

\begin{definition}
	Here are some notation conventions: 
	\op 
	\im Let $ \PP_w$ denote the probability measure true in $w$ if $w$ contains such a thing. 
	\\[-0.8em]
	\im Let $h^{M, n, w}$ denote the hypothesis or output of method $M$ at stage $n$ in world $w$, that is, $M(e_1 e_ 2 \cdots e_n)$, where $e_1 e_ 2 \cdots e_n$ is the sequence of the first $n$ data points received in world $w$. 
	\\[-0.8em]
	\im Similarly, let $\hat{h}^{M, n}$ be the random variable that maps each possible data sequence of length $n$ to the output of $M$ on that sequence; so $\hat{h}^{M, n}$ can be intuitively understood to denote the random hypothesis or output of method $M$ at stage $n$ $($leaving worlds and data sequences unspecified$)$, following the standard use of variables in statistics.
	\ed 
\end{definition}

\noindent The random variable notation $\hat{h}^{M, n}$ is particularly convenient for expressing probabilities like the following: 
	\begin{eqnarray*}
	 \PP_w \Big[\, \LL \big( \hat{h}^{M, n}, w \big) < \epsilon \Big] &\;=_\mathrm{df}\;&
	\PP_w \Big\{\, e_1 e_ 2 \cdots e_n: \LL \big( M(e_1 e_ 2 \cdots e_n), w \big) < \epsilon \Big\} \,.
	\end{eqnarray*}
The right side means the probability, in world $w$, of observing a data sequence $e_1 e_ 2 \cdots e_n$ of length $n$ such that the loss of $M$'s output is below the threshold $\epsilon$. This phrase can be more concisely expressed using the notation on the left side: it denotes the probability, in world $w$, for the loss of $M$'s output to be bellow $\epsilon$ given sample size $n$. 

Here is the real thing:

\begin{definition}[Three Basic Modes of Convergence] An inference method $M$ for a problem $\PPP$ can be said to achieve one or another mode of convergence to the truth. The following defines three modes. 

	\op 
	\im {\bf Convergence with Nonstochastic Identification} 
	\\[0.2cm] for any world $w \in \WW$, 
	\\ there exists sample size $N$ such that, 
	\\ for all $n \ge N$,	\\[-0.2cm]
		\op 
		\xm $\LL \big( h^{M, n, w}, w\big) = 0$;
		\\[-0.3cm]
		\xm that is, given sample size $n$, $M$ outputs exactly the truth in $w$.	\\[-0.2cm]
		\ed 

	\im {\bf Convergence with Stochastic Identification:} 
	\\[0.2cm] for any probability threshold $1-\delta < 1$, 
	\\for any world $w \in \WW$, 
	\\ there exists sample size $N$ such that, 
	\\ for all $n \ge N$, \\[-0.2cm]
		\op 
		\xm $\PP_w$ exists, and
		\xm $\PP_w \big[\, \LL \big( \hat{h}^{M, n}, w \big) \,=\, 0 \,\big] \;>\; 1 - \delta$;	\\[-0.2cm]
		\xm that is, given sample size $n$, the probability for $M$ to output exactly the truth is high $($greater than $1 - \delta$$)$ in $w$.	\\[-0.2cm]
		\ed 
	
	\im {\bf Convergence with Stochastic Approximation} 
	\\[0.2cm] for any upper bound on loss $\epsilon > 0$,
	\\for any probability threshold $1-\delta < 1$, 
	\\for any world $w \in \WW$, 
	\\ there exists sample size $N$ such that,
	\\ for all $n \ge N$, \\[-0.2cm]
		\op 
		\xm $\PP_w$ exists, and
		\xm $\PP_w \big[\, \LL \big( \hat{h}^{M, n}, w \big) \,<\, \epsilon \,\big] \;>\; 1 - \delta$;	\\[-0.25cm]
		\xm that is, given sample size $n$, the probability that $M$ outputs a hypothesis $\epsilon$-close to the truth is high $($greater than $1 - \delta$$)$ in $w$.	
		\ed 
	\ed 
\end{definition}

These three basic modes of convergence have been mostly developed and studied separately in different areas, couched in very different languages, and employed to talk about apparently different subjects. The formalism developed here provides a uniform reformulation of those modes. Moreover, the plain English glosses accompanying the above reformulated definitions (the clauses following `that is') make it clear that the old, familiar modes of convergence are covered:   
\op 
	\item[(i)] {\bf Convergence w/ Nonstochastic Identification}: It goes by the name {\em identification/decidability in the limit} in formal learning theory (Kelly 1996: ch. 3), originally designed for theory choice in a deterministic setting.
	\\[-0.6em]
	\item[(ii)] {\bf Convergence w/ Stochastic Identification}: This captures the conjunction of {\em consistency in significance level} and {\em consistency in power} as studied in statistical hypothesis testing (Lehmann 1999: ch. 3). It also captures the so-called {\em model selection consistency} in statistical model selection (Claeskens et al. 2008: ch. 4).
	\\[-0.6em]
	\item[(iii)] {\bf Convergence w/ Stochastic Approximation}: This captures so-called {\em estimation consistency} in the statistical theory of estimation (Lehmann 1999: ch. 2). As you will see below (section \ref{sec-supervised}), it also captures the consistency of learning algorithms in supervised learning, which includes classification (Shalev-Shwartz et al. 2014: Part I) and nonparametric regression (Gy\"{o}rfi et al. 2002: ch. 1).
	\ed 

It is not hard to develop variants of the above modes of convergence. Let me briefly outline some notable ones. For each of the three modes defined above, exchanging the quantifiers `for any world' and `there exists sample size' gives us a higher standard, a mode of {\em uniform} convergence. There are other variants, which can be obtained by thinking about: {\em rates} of convergence (Gy\"{o}rfi et al. 2002: ch. 1), {\em monotonic} convergence or somewhat {\em stable} convergence if not perfectly monotonic (Lin 2022), and convergence for {\em almost all} worlds in $\WW$---almost all in a topological sense (Lin 2019). Those modes of convergence and their combinations have been studied in one or another area. From an epistemological point of view, they correspond to higher or lower standards for evaluating inference methods. However, I will focus on the three basic modes defined above, which suffice for making the philosophical point I want to make below. 

\section{Towards Unification}\label{sec-division}

The three basic modes of convergence have been reformulated in a uniform notation to clarify their close connection. Although they are developed in distinct areas, they set standards for assessing {\em all} inference methods in {\em any} empirical problems. The standards range from high to low---from actually getting the truth, to probably getting exactly the truth, to probably getting close to the truth. Hence the hierarchy depicted in Table \ref{tab-basic}.
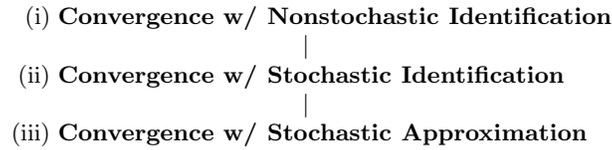
\begin{table}[h]
\centering
\caption{\em The Hierarchy of the Three Basic Modes of Convergence}\label{tab-basic}
\begin{tabular}{rl}
(i) & {\bf Convergence w/ Nonstochastic Identification} \\
& \quad\quad\quad\quad\quad\quad\quad\quad\quad\quad $\vert$ \\
(ii) & {\bf Convergence w/ Stochastic Identification} \\
& \quad\quad\quad\quad\quad\quad\quad\quad\quad\quad $\vert$ \\
(iii) & {\bf Convergence w/ Stochastic Approximation} \\
\end{tabular}
\end{table}

We need just one more definition before the first result can be stated:

\begin{definition}[Achievability]
	A mode of convergence is said to be achievable for an empirical problem if some inference method for that problem satisfies that mode of convergence. 
\end{definition}

\noindent Then we have:

\begin{proposition}\label{pro-division}
Consider the hierarchy of the three modes of convergence depicted in Table {\em \ref{tab-basic}}. 
\op 
\im For the easy raven problem, the highest achievable is mode {\em (i)}. 
\im For the fair coin problem, it is mode {\em (ii)}. 
\im For the coin bias problem, it is mode {\em (iii)}.
\ed 
\end{proposition}
 
See the appendix for the proof. The novelty of this result consists in what is {\em not} achievable. The claims of achievability are already obtained with greater generality in formal learning theory and statistics. But I still provide elementary proofs of those achievability claims, considering that readers familiar with one of the two areas might not be so with the other. 

Since the novelty is in the claims of {\em un}achievability, let me explain the proof strategy. Mode (i) is unachievable for the two statistical problems due to a specific kind of {\em underdetermination by data}: those two problems allow that one and the same infinite data sequence can be generated under different hypotheses. Mode (ii) is unachievable for the coin bias problem because of considerations about {\em cardinality}: the number of competing hypotheses (all the real numbers in the unit interval) is strictly greater than the number of possible evidential inputs (all finite binary sequences). So, for any inference method, there is at least one hypothesis doomed to be never an output. 

I suspect that the proof strategy just sketched allows us to generalize the result to cover wide classes of empirical problems. However, I will focus on the three paradigm problems---easy raven, fair coin, and coin bias---to streamline the discussion and emphasize more pressing points: a picture of different areas unified into a cohesive whole. 

Although statistics and formal learning theory may appear to be very different, they operate with the same guiding principle in the unifying picture I propose: 
	\begin{center}
		\fbox{Strive for the Highest Achievable!}
	\end{center}
Statisticians design inference methods with a minimum qualification in mind, which is standardly called consistency in statistics but is just the stochastic mode (ii) or (iii) defined above. Statisticians do not aim at the higher mode (i). The {\em cause} underlying statisticians' practices may lie in how their textbooks are written and their PhD programs structured. But the {\em reason} that justifies what statisticians do, I propose, is based on this fact: the empirical problems addressed by statisticians are too hard to make the higher mode (i) achievable. Similarly, formal learning theorists do not aim at the lower modes (ii) and (iii), because they should not: since the empirical problems they study make the higher mode (i) achievable, they should use that higher standard to evaluate inference methods. 

So, statistics and formal learning theory need not be regarded as two separate areas that {\em assume} different standards to evaluate different kinds of inference methods. Instead, there is a unifying principle: strive for the highest achievable. The use of different standards in different areas---stochastic vs. nonstochastic---need not be assumed and can be {\em justified} by the proposed principle. 

Similarly, hypothesis testing and parameter estimation need not be viewed as distinct subareas of statistics with separate evaluative standards. The use of different standards---identification vs. approximation---need not be assumed and can be {\em justified} by the proposed principle.

% Different areas study different types of empirical problems and different types of inference methods, apparently assuming different evaluative standards. But only apparently so. A unifying principle exists to organize them. The rest is a matter of division of labor, not separation of disciplines.

% This new way of structuring different disciplines raises an interesting open question. Looking at the three basic modes of convergence, there clearly is an omission: convergence with nonstochastic approximation, which is not hard to define by combining some clauses already formulated above. This new mode should sit below (i) and above (ii). Is this new mode the highest achievable for at least some interesting empirical problems? I leave this question for future research. 

%\begin{conjecture} The previous proposition still holds even if the problem of what's the bias of the coin comes with a stronger background assumption: that the true bias of the coin is a rational number. \end{conjecture} \begin{proof} By diagonalization, and the denseness of rational numbers.\end{proof}

%%% estimation theory pursues more: efficiency, unbiasedness???

\section{Deterministic vs. Stochastic?}\label{sec-det-sto}

%%% Let me address a worry that might still linger: Mode of convergence (i) seems to concern a very unique topic, different from the topic of modes (ii) and (iii). See: the raven problem, which achieves the so-called higher mode (i), fails to achieve the so-called lower modes (ii) and (iii) for a trivial reason: both (ii) and (iii) require that probabilities exist in the worlds on the table (in $\WW$), while probabilities are missing from the worlds in the raven problem. So it seems that mode (i) and formal learning theory are here specially for studying empirical problems that entertain deterministic hypotheses. The other two modes and statistics are here specifically for an entirely different class of hypotheses: statistical/stochastic hypotheses. Or so the worry goes.

Although formal learning theory is often presented in a way that gives the impression that it only concerns deterministic hypotheses (as in Kelly 1996), let's be more careful. The hypothesis ``all ravens are black'' is neutral regarding whether the actual world is deterministic or indeterministic. This hypothesis can be true in a deterministic world, but it can also be true in an indeterministic world---a world in which all ravens turn out to be black by chance. We have represented this world without probabilities by:
$$
w \;= \; \left( \texttt{111}\cdots, \yes\right) .
$$
Strictly speaking, this $w$ only represents a very coarse-grained world, for it says nothing whatsoever about whether determinism is true or false. The coarse-grained world $w$ can be realized by two different, more fine-grained worlds:
\begin{eqnarray*}
	u &\;=\;&  \left( \texttt{111}\cdots, \yes, \PP_\mathrm{trivial}\right), 
	\\
	u' &\;=\;&  \left( \texttt{111}\cdots, \yes, \PP'\right), 	
\end{eqnarray*} 
where, $\PP_\mathrm{trivial}$ is a probability function that only assigns trivial probabilities $0$ and $1$---in particular, it assigns probability $1$ to the constant sequence $\texttt{111}\cdots$. But $\PP'$ is a nontrivial probability function, assigning (say) $.7$, to the constant sequence $\texttt{111}\cdots$. In both of those fine-grained worlds, it is true that all ravens are black. However, it is true deterministically in the first world $u$, and true only by chance in the second world $u'$. %%% The {\em content} of the background assumption of the easy raven problem does not preclude any of those two worlds $u$ and $u'$. If we are not careful, it can be misleading to represent that background assumption as a set of ``probability-free'' worlds like $w$ rather than worlds like $u$ and $u'$. With probabilities made explicit, we can represent deterministic worlds such as $u$ with trivial probabilities, and indeterministic worlds such as $u'$ with nontrivial probabilities. 
Hence the following definition:

\begin{definition}[Probabilistic Extension]
	Suppose that an evidence tree $\EE$ is given. Consider a world $w = (e_1 e_2 \cdots, h)$ that contains no probability measure. This world $w$ is said to have another world $w'$ as a probabilistic extension if $w'$ takes the form $w' =  (e_1 e_2 \cdots, h, \PP)$, which differs from $w$ only by adding a $($trivial or nontrivial$)$ probability measure $\PP$ over the evidence tree $\EE$.
\end{definition}

\begin{definition}[Fine-Grained Version]
	Fine-grained versions of a problem $\PPP$ are problems of the form: $(\HH, \EE, \WW', \LL')$, with the same $\HH$, the same $\EE$, but a different $\WW'$ and a different $\LL'$, satisfying the following constraints: 
	\op 
	\im $\WW'$ can be obtained from the original $\WW$ by, first, removing each world $w$ that contains no probability measure and, then, replacing $w$ with one or multiple probabilistic extensions of $w$. 
	\\[-0.8em]
	\im $\LL'(h, w') = \LL(h, w)$ if $w'$ is a probabilistic extension of $w$.
	\ed 
\end{definition} 

\noindent Then we have:

\begin{proposition}\label{pro-fine}
	Consider the three modes of convergence in Table {\em \ref{tab-basic}}. If the highest mode {\em (i)} is achievable for a problem $\mathcal{P}$, then the three modes {\em (i)}-{\em (iii)} are all achievable for any fine-grained version of $\mathcal{P}$ that only involve countably additive probability measures.
\end{proposition}

See the appendix for the proof.

\begin{corollary}\label{cor-chain}
	Mode {\em (i)} implies mode {\em (ii)} in the precise sense described by the preceding proposition. Mode {\em (ii)} in turn implies mode {\em (iii)} in the standard sense.
\end{corollary}

The easy raven problem fails to make (ii) and (iii) achievable only because of its somewhat misleading mathematical representation. The fine-grained versions of that problem do a better representational job, being explicitly neutral between determinism and indeterminism, making all the three modes provably achievable. 

So it is misleading to say that formal learning theory differs from statistics in that the former presupposes determinism. This claim is incorrect; there is no such presupposition. The distinction between the two areas is best understood as a division of labor guided by a common principle, as explained earlier.

%%%So, it makes little sense to say that modes of convergence (i)-(iii) concern different subjects such as deterministic vs. stochastic hypotheses. The three modes of convergence differ primarily in that they set high or low standards for evaluating inference methods. And, to reiterate, the corresponding three areas---formal learning theory, statistical testing theory, and statistical estimation theory---are unified under the unifying principle of striving for the highest achievable. It is a matter of division of labor, rather than separation. 

%%%It should be noted that the proof of the above result relies essentially on the axioms of {\em countable additivity} for probability measures. There are philosophical concerns about countable additivity. But I think I can afford to set them aside in this paper. For most of the applications in statistics I am aware of, countable additivity {\em is} assumed. I would be happy if the picture of inductive logic developed here is general enough to cover most of statistics.

\section{Supervised Learning}\label{sec-supervised}

The present setting also covers problems studied in supervised learning, whose simplest examples concern binary classification:

\begin{example}[Binary Classification]\label{ex-classification}
	Suppose that we want to determine whether a given object is a cat by examining a (pixelated) picture of it. Or suppose that we want to determine whether a given watermelon is tasty by examining properties such as its skin color distribution and the sound it produces when tapped. Or suppose, in general, that we want to classify an object of a certain kind into category 0 or category 1, and do it on the basis of the feature that it has in a countable set of mutually exclusive features: $\XX = \{x_1, x_2, x_3\ldots\}$.
	A {\bf (binary) classifier} is an indicator function $h: \XX \to \{0, 1\}$. Suppose, further, that we are given a set $\HH$ of (some or all) such classifiers.
	We would like to pick a good classifier from $\HH$ in light of examples. An {\bf example} is an object with a specified feature $x \in \XX$ and a specified category $0$ or $1$; so an example is formally an ordered pair $(x, 0) \textrm{ or } (x, 1)$. Such examples form an {\bf example space}: $\XX \times \{0, 1\}$. A data sequence $(\data)$ is a finite sequence of such ordered pairs. All such data sequences form an evidence tree $\EE = (\XX \times \{0, 1\} )^{< \infty}$. A {\bf learning algorithm} is just an inference method $M: \EE \to \HH \cup \{\duno\}$, although people in machine learning typically only consider $M: \EE \to \HH$. To sum up: a {\bf task of binary classification} can be formally identified as an ordered pair $(\XX, \HH)$ consisting of two elements: 
	\op
	\im a countable feature space $\XX = \{x_1, x_2, x_3\ldots\}$, 
	%\\[-1.2em]
	\im a set $\HH \subseteq \{0, 1\}^\XX$, as a candidate pool of classifiers.
	\ed  
Those two elements suffice to determine the other elements of such a task, including classifiers, examples, an evidence tree $\EE$, and learning algorithms. {\em But what counts as a good classifier---good for predictive purposes?} The quality of a classifier depends on the state of the world. Assume that examples are generated by a probabilistic mechanism, represented by a probability distribution $D$ over the example space $\XX \times \{0, 1\}$, where $D(x, y)$ denotes the probability that the next example has feature $x$ and belongs to category $y$. The {\bf predictive risk} of a classifier $h$ with respect to the (true but unknown) distribution $D$ is given by the probability of {\em misclassification}: 
	$$
	\mathsf{Risk}(h, D)
	\;=\; D \big\{ (x, y): h(x) \neq y \big\}
	\;=\; \!\!\!\sum_{(x, y): \, h(x) \neq y} D(x, y)
	$$
The smaller, the better. All this points to an empirical problem $\PPP$, called a {\bf (binary) classification problem}: 
\op 
\im Question: Which classifier is the best in class (in $\HH$) for predictive purposes? So the hypothesis space is $\HH$, the given candidate pool of classifiers. 
\im $\EE$, the evidence tree, is the tree of sequences of examples $(x, y)$ taken from the example space $\XX \times \{0, 1\}$.	
\im $\WW$, which represents the background assumption,  is the set of all possible worlds of the form $w = (e_1 e_2 \cdots, D, \PP_D)$, where $D$ is an arbitrary probability distribution over the example space, and $\PP_D$ is the IID probability measure (on $\EE$) generated from distribution $D$. 
\im $\LL(h, w) 	= \mathsf{Risk}(h, D_w) - \min_{h' \in {\cal H}} \mathsf{Risk}(h', D_w)$, where $D_w$ denotes the distribution in world $w$	
\ed 
This concludes the last and longest of the four examples examined in this paper. It is no coincidence that textbooks in machine learning use the symbol $h$ to denote classifiers, referring to them as hypotheses (Shalev-Shwartz et al. 2014).
\end{example}
%%%   The classifier $h$ that attains zero loss is the true answer to the question posed: the question of which is the best in class for predictive purposes. When our guess $h$ is a false answer, we still hope that $h$ comes close to the true answer---that is, have a small loss in predictive risk. 

Once we see what binary classification is, generalizations are straightforward. When the category set $\YY = \{0, 1\}$ is generalized to a set of finite categories, $\YY = \{1, 2, \ldots, k\}$, we obtain problems of {\em multiclass classification}. When both the feature space $\XX$ and the category space $\YY$ become continuous, say Euclidean spaces $\mathbb{R}^n$, we have problems of {\em nonparametric regression}. These generalizations encompass nearly the entire area of supervised learning. 

In supervised learning, the minimum qualification for good learning algorithms is called {\em consistency}, which is essentially the mode of convergence (iii) as defined in this paper---convergence with stochastic approximation. However, consistency does not need to be {\em assumed} as a minimum qualification in supervised learning. It can be justified as follows: By the principle of striving for the highest achievable, the achievability of at least mode (iii) in the hierarchy in Table \ref{tab-basic} implies that mode (iii) or a higher mode has to be achieved, which in turn implies, thanks to corollary \ref{cor-chain}, that mode (iii) has to be achieved at the very least---as a minimum qualification. The same argument also justifies mode (iii) as a minimum qualification for good estimators in statistical estimation.

%%% In supervised learning, there can be an easy problem for which it is possible to achieve mode (ii), convergence with stochastic identification. This is often so when the class of classifiers under consideration $\HH$ is finite. But when the $\HH$ is infinite, ... 

%\section{Some Myths that Divide Disciplines}
% Formal Learning Theory: just for deterministic hypothesis?
% Machine Learning: Not care about truth?
% Learning Theories: focus on computability and complexity?
% Statistics: Not care about learnability?
% Stay flexible here, for you might run out of space!

\section{Closing}

This paper reformulates three basic modes of convergence to the truth using a uniform notation, and explores their potential to create a unified picture of inductive logic. Additional modes of convergence (as mentioned at the end of section \ref{sec-modes}) should be investigated to enrich the simple hierarchy in Table \ref{tab-basic} and to determine whether more areas can be incorporated into the unified framework. In this closing note, let me explain how we arrived at this point and where we might go from here.

In hindsight, Peirce already did a lot for us over a century ago. As pointed out in Lin's (forthcoming) discussion of Peirce's convergentism, Peirce sought to justify enumerative induction by appeal to the mode of convergence with nonstochastic identification (Peirce 1994: CP 2.775, 7.125), and he {\em also} studied statistical estimation using the mode of convergence with stochastic approximation (Peirce 1994: CP 2.669-93). Unfortunate, after Peirce, those two modes of convergence took diverging paths. The stochastic one helped establish the estimation theory in statistics (Fisher 1925), and the nonstochastic one helped create formal learning theory (Putnam 1965, Gold 1967)---but those two areas have long been regarded as largely unrelated. I propose that we return to Peirce's idea and reinforce it with the principle developed above ``Strive for the Highest Achievable!''

The result is a conception of logic that unifies formal learning theory, statistics, and a significant part of machine learning: supervised learning. But can it be extended to cover other areas of machine learning, such as reinforcement learning and unsupervised learning? I am optimistic for reinforcement learning, for it has been given a largely uniform foundation as supervised learning (Mohri et al. 2018). Of course, a detailed argument is needed nonetheless, but that has to be left to future work. 

However, the prospect for incorporating unsupervised learning into the unified picture remains unclear. In fact, even theorists of machine learning have difficulty evaluating algorithms of unsupervised learning by a rigorous standard---let alone a standard defined as a mode of convergence. To talk about convergence to the truth, there needs to be a truth to begin with, but such a truth is typically missing in unsupervised learning, as observed in a standard textbook on the theoretical foundation of machine learning: ``[A] basic problem is the lack of ``ground truth'' for clustering, which is a common problem in unsupervised learning'' (Shalev-Shwartz \& Ben-David 2014: 308).

The most influential approach to inductive logic to date in philosophy is the Bayesian one. I suspect that it can be developed in a manner that fits into the unified picture I have painted. On the Bayesian approach, inductive logic should be founded on the idea that probabilities as rational degrees of belief ought to be updated by conditionalization on the new evidence (Carnap 1945). But is it that all priors---initial assignments of probabilistic degrees of belief---are equally permissible? The answer is ``yes'' according to the {\em subjectivists} in Bayesian epistemology. But the answer is ``no'' according to anti-subjectivists, including the {\em objectivists}, who maintain that permissible priors must be ``flat'', conforming to the principle of indifference or a variant of it. But the objectivists are not the only people who would like to constraint the candidate pool of permissible prior by something more than the axioms of probability. For a group of Bayesians who are less known to the philosophical community but work in a branch of statistics called {\em Bayesian nonparametric statistics}, it is customary to rule out some prior assignments of degrees of belief by appeal to considerations about convergence (see Rousseau 2016 for a review). The idea is simple: a Bayesian prior is permissible only if updating it successively via conditionalization ensures a sequence of posteriors that achieves a specific mode of convergence to the truth, commonly called {\em Bayesian consistency}. Thus, even Bayesians may find their place within the unifying framework I have outlined for inductive logic; however, the details remain to be addressed in future work.

I hope all this offers initial reasons to be optimistic about the prospect of a unified inductive logic.

\section*{Acknowledgements}

I thank Konstantin Genin for his patient and inspiring discussions. I also thank two anonymous reviewers for their detailed and constructive comments on earlier versions of this paper. 

\appendix

\section{Appendix: Proofs}

\subsection{Proof of Proposition \ref{pro-division}}

	To establish the first part for the easy raven problem, it suffices to verify that mode (i) is achieved in the easy raven problem by the inference method that outputs $\yes$ exactly when the input contains no $\mathtt{0}$, and outputs $\no$ otherwise. 
	
	Now, establish the second part for the fair coin problem as follows. Mode (i), convergence with nonstochastic identification, is unachievable because the fair coin problem features this kind of underdetermination by data: The background assumption $\WW$ in the fair coin problem is so weak that it contains a pair of worlds in $\WW$ sharing the {\em same} indefinite data sequence (say, heads, tails, heads, tails, and so on), while one world makes $\texttt{Fair}$ true and the other makes $\texttt{Unfair}$ true. So, to converge to the truth in one of those two worlds is to converge to a falsehood in the other. Mode (i) is thus unachievable. As to the achievability of mode (ii) for the fair coin problem, it can be proved by Bernoulli's law of large numbers as follows. Let $M$ be the inference method that outputs $\texttt{Fair}$ exactly when the observed frequency $\bar{X}_n$ of heads in the sample is {\em close enough} to $0.5$ in the sense that $|\bar{X}_n - 0.5| < 1/{\sqrt[4]n}$, where $n$ is the sample size. Now recall Bernoulli's law of large numbers: 
		$$\PP_{\theta}( |\bar{X}_n - \theta| < \epsilon) \;\;\ge\;\; 1 - \frac{1}{4n\epsilon^2} \,.$$ 
		In any world in which the truth is $\texttt{Fair}$, the probability for $M$ to get the truth is obtained by letting $\theta = 0.5$ and $\epsilon = 1/{\sqrt[4]n}$ in Bernoulli's law of large numbers: 
		\begin{eqnarray*}
		&& \PP_{.5}\left( |\bar{X}_n - 0.5| \,<\, 1/{\sqrt[4]n}\right) 
		\\ 
		&\ge \;& 1 - \frac{1}{4n(1/{\sqrt[4]n})^2} 
		\\
		&= \;& 1-\frac{1}{4\sqrt{n}} \,,
		\end{eqnarray*}
	which converges to $1$ as $n$ tends to infinity. Moreover, in any world where the truth is $\texttt{Unfair}$ with true bias $\theta \neq 0.5$, when the sample size $n$ is large enough to ensure that $1/{\sqrt[4]n} \,<\, \frac{1}{2}|\theta - 0.5|$, the probability for $M$ to get the truth is
		\begin{eqnarray*}
		&& \PP_{\theta}\left( |\bar{X}_n - 0.5|  \,\ge\, 1/{\sqrt[4]n}\right)
		\\
		&\ge \;& \PP_{\theta}\left( |\bar{X}_n - \theta|  \,<\, 1/{\sqrt[4]n}\right)
		\\
		&\ge \;& 1-\frac{1}{4\sqrt{n}} \,,
		\end{eqnarray*}
	which converges to $1$ as $n$ tends to infinity. So $M$ achieves mode (ii).
 
 	Now, establish the third part for the coin bias problem as follows. The set of all possible (evidential) inputs is the set of all nodes in the binary tree. So there are countably many possible inputs. But there are uncountably many hypotheses. Let $M$ be an arbitrary inference method. It follows that there is some hypothesis $h_M$ that $M$ is doomed to never output in any world. So, the probability for $M$ to output (exactly) the truth always stays zero in the worlds in which the truth is $h_M$. So $M$ fails to achieve mode (ii). But mode (iii) is achievable, for a well-known reason: the inference method that always outputs the observed frequency of heads achieves mode (iii), convergence with stochastic approximation, thanks to Bernoulli's law of large numbers (again).
Q.E.D.

\subsection{Proof of Proposition \ref{pro-fine}}

	Suppose that mode (i) is achievable for a problem $\PPP$, which has a refined version $(\HH, \EE, \WW', \LL')$. Let $M$ be an inference method for the original problem that achieves mode (i), which implies that there is a one-to-one correspondence between the worlds in $\WW$ and the infinite branches of the evidence tree $\EE$: 
	$$
		w = (\stream, -) \quad \mapsto \quad \stream \,\in\, \EE\,.
	$$ 
	So, any probability measure defined over $\EE$ can be equivalently construed as a probability measure defined over $\WW$. Let ${\sf Success}(M, n)$ be the set of the worlds $w \in \WW$, or equivalently the branches $b \in \EE$, such that, by stage $n$, $M$ has output the truth and would never drop it in $b$. By definition, ${\sf Success}$ is monotonic in this sense: whenever $n \le n'$,
	$$
	{\sf Success}(M, n) \; \subseteq \; {\sf Success}(M, n') \,.
	$$
	Since $M$ achieves mode of convergence (i), it follows that 
		$$\textstyle{\bigcup_{n \in \NN} {\sf Success}(M, n)} \;=\; \EE\,.$$ 
	Let $w'$ be an arbitrary world in $\WW'$, and let $\PP_{w'}$ denote be the probability measure over $\EE$ in $w'$. Then, by countable additivity, we have:
	\begin{eqnarray*}
	&& \lim_{n\to\infty} \PP_{w'} \big({\sf Success}(n)\big)
	\\
	&=\;& \textstyle{\PP_{w'} \big(\bigcup_{n \in \NN} {\sf Success}(n)\big)}
	\\
	&=\;& \PP_{w'} (\EE)
	\\
	&=\;& 1
	\end{eqnarray*}
	Now, note that the probability for $M$ to get the truth by stage $n$ in world $w'$ is greater than or equal to $\PP_{w'}\big({\sf Success}(n)\big)$, which approaches $1$ as $n \to \infty$ thanks to the above calculation. So $M$ achieves mode (ii), convergence with stochastic identification. Achieving mode (ii) immediately implies achieving mode (iii). 
Q.E.D.

\end{document}